\documentclass{KapProc} 
\let\footnote\savefootnote
\let\footnotetext\savefootnotetext
\let\footnoterule\savefootnoterule
\normallatexbib

\begin{document}

\articletitle{Gravitational Energy-Momentum}

\articlesubtitle{in the Tetrad and Quadratic Spinor
Representations of General Relativity ${}^\dagger$}

\author{Roh S. Tung}
\affil{California Institute for Physics and Astrophysics\\
         366 Cambridge Avenue, Palo Alto, California 94306, USA\\
         and\\
         Enrico Fermi Institute, University of Chicago\\
         5640 South Ellis Avenue, Chicago, Illinois 60637, USA}
 \email{tung@calphysics.org}

\author{James M. Nester}

\affil{Department of Physics and Center for Complex Systems \\
     National Central University,  Chungli 320, Taiwan, ROC}
 \email{nester@phy.ncu.edu.tw}

\begin{keywords}
Gravitational Energy-Momentum, Quadratic Spinor Representation of
General Relativity
\end{keywords}

\begin{abstract}
In the Tetrad Representation of General Relativity, the
energy-momentum expression, found by M{\o}ller in 1961, is a
tensor wrt coordinate transformations but is not a tensor wrt
local Lorentz frame rotations. This local Lorentz freedom is shown
to be the same as the six parameter normalized spinor degrees of
freedom in the Quadratic Spinor Representation of General
Relativity. From the viewpoint of a gravitational field theory in
flat space-time, these extra spinor degrees of freedom allow us to
obtain a local energy-momentum density which is a true tensor over
both coordinate and local Lorentz frame rotations. \\
\\
gr-qc/0010001
\end{abstract}

\renewcommand{\thefootnote}{\fnsymbol{footnote}}
\footnotetext[2]{in ``Gravitation and Cosmology: From the Hubble
Radius to the Planck Scale. A Symposium in Celebration of the 80th
Birthday of Jean-Pierre Vigier (August 21 - 25, 2000, University
of California Berkeley), R. L. Amoroso, G. Hunter, M. Kafatos and
J. P. Vigier eds., Kluwer Academic Publishers, 2001'', to be
published.}
\renewcommand{\thefootnote}{\arabic{footnote}}

\section*{Introduction}

Conservation of energy-momentum, which is associated with
spacetime symmetry, plays an important role in physics. When we
trace back the history, we find that a new physics has usually
been born with a violation of conservation of energy-momentum.
Perhaps the only exception was Einstein's radical idea: general
relativity --- a theory of {\em spacetime} itself.

The problem of determining the gravitational energy-momentum arose
immediately after Einstein's formulation in 1915; attempts looking
for a local energy-momentum only resulted in a set of {\em
pseudotensors}. After much effort, people concluded that there is
no proper physical local energy-momentum density for the
gravitational field. The situation was gradually clarified to the
following conclusions: (i) The energy-momentum concept in a
gravitational field can be introduced if we replace the spacetime
symmetry in ordinary relativistic field theory by the concept of
asymptotic flatness, i.e. {\em total} energy is well-defined. (ii)
Because of the equivalence principle, gravitational
energy-momentum is not localized. As the famous textbook of
Misner, Thorne and Wheeler teaches \cite{MTW}:
\begin{quote}
``anybody who looks for a magic formula for {\em local
gravitational energy-momentum} is looking for the right answer to
the wrong question ''.\footnote{In order to be a good student, we
should probably not ask such a ``wrong question''. But in this
paper, we are being naughty students.}
\end{quote}

Newtonian gravity theory is based on action-at-a-distance, so in
that case we expect only a total energy for a gravitating system.
But for relativistic gravity theories we expect meaningful local
quantities since the interactions are local and they {\em
exchange} energy-momentum locally.

Roger Penrose was not satisfied with only the total energy for
gravitation being defined; he stated
\begin{quote}
``It is perhaps ironic that {\em energy conservation},  a
paradigmatic physical concept arising initially from Galileo's
(1638) studies of the motion of bodies under gravity, and which
now has found expression in the (covariant) equation $ \nabla_a
T^{ab}=0 $
--- a cornerstone of Einstein's (1915) general relativity ---
should nevertheless have found no universally applicable
formulation, within Einstein's theory, incorporating the energy of
gravity itself.''
\end{quote}
and then proposed the idea of {\em quasilocal} (i.e. associated
with a closed 2-surface) energy-momentum \cite{Penrose}. There
have been several proposals (an extensive literature was given in
Ref.1 of \cite{BY} ) for quasi-local energy-momentum. They need
either a reference background \cite{BY,CNT} or a globally defined
spinor field \cite{DM}.

The meaning for a reference background was in fact pointed out by
Poincar\'e \cite{Poincare,Wiesendanger}---that the physical
description is often based on {\em a priori} conventions. For
spacetime geometry, two points of view are possible. (i) According
to general relativity, the line element between neighboring events
is measured by using rods or clocks with the same length or rate
which are independent of the field present. The resulting
spacetime is curved in general. (ii) On the other hand, one can
define the line element to be of Minkowskian form. Accordingly,
the rods and clocks are affected by the gravitational field. Apart
from global topological questions the two complementary points of
view are equivalent. The origin of Weyl's {\em gauge} idea is in
fact to abandon the idea of adding lengths in general relativity,
in Weyl's opinion, keeping a rod to have the same length is a
concept which involves action-at-a-distance \cite{GronHehl}.  The
same applies to clocks.

Microscopically, it is very difficult to have classical concepts
such as rods and clocks giving a simple microscopic understanding
of gravitation. Hence the main effort of current quantum gravity
is to find new concepts for the space-time at the Planck scale,
there have been many stimulating ideas, e.g., strings (p-branes),
twistors, or loops. These concepts provide a rich structure for
space-time at the Planck scale. From the field theory point of
view, microscopically space-time geometry enters only as a
background concept necessary to defining a field theory. The
search for a good expression for local energy-momentum is thus
especially important since the energy concept is associated with
the fundamental structure of space-time.

Since for Einstein's general relativity, there is no space-time
symmetry in general, we do not expect a local conserved
energy-momentum. In this case, we expect at most a quasi-local
definition. However, considering Einstein gravity as an ordinary
field theory in flat Minkowski background space-time, we do have a
Minkowski space-time symmetry. Can we then obtain an
energy-momentum density tensor in this case? In this paper we will
obtain such a quantity by a change of variables and by adding
extra spinor gauge variables.

\section*{Metric Representation}

The Hilbert Lagrangian density for General Relativity is ${\cal
L}_H=-\sqrt{-g} R$. The traditional approach uses the metric
coefficients in a coordinate basis as the dynamic variables, so
${\cal L}_H={\cal L}_H (g,\partial g, \partial\partial g)$.
Because of the second derivatives, this is not suitable for
getting an energy-momentum density.  However a certain
(noncovariant) divergence can be removed (without affecting the
equations of motion) leading to Einstein's Lagrangian ${\cal
L}_E={\cal L}_E(g,\partial g)={\cal L}_H-\hbox{div}$. One can now
apply the standard procedure and get the canonical energy-momentum
density.  It is known as the Einstein {\em pseudotensor}; its
value depends to a large extent on the coordinate (``gauge'')
choice.  No satisfying technique has been found to separate the
``physics'' from the coordinate gauge.

\section*{Tetrad Representation}

An alternative is to use an orthonormal frame (tetrad), a pioneer
of this approach was M{\o}ller \cite{Mo61}. Let
$g_{\mu\nu}=g_{ab}e^a{}_\mu e^b{}_\nu$, with
$g_{ab}=\hbox{diag}(+1,-1,-1,-1)$, and regard the Einstein-Hilbert
Lagrangian as a function ${\cal L}_e(e,\partial e,\partial\partial
e)$ of the tetrad $e^a{}_\mu$. A suitable total divergence can
again be removed yielding a Lagrangian density which is first
order in the derivatives of the frame.   Now there is an
associated energy-momentum density which {\em is} a tensor
(density) under coordinate transformations, {\em but} it depends
on the choice of orthonormal frame (Lorentz gauge)
\cite{Mo61,Goldberg,Pereira00}.

\section*{Quadratic Spinor Representation}

Let $\psi$ be any Dirac spinor field and let $\Psi=\vartheta
\psi$, where $\vartheta=\vartheta^a \gamma_a=e^a{}_\mu\gamma_a
dx^\mu $ is a Clifford algebra 1-form.\footnote{ Our Dirac matrix
conventions are $\gamma_{(a}\gamma_{b)}=g_{ab}$,
$\gamma_{ab}:=\gamma_{[a}\gamma_{b]}$,
$\gamma_5:=\gamma_0\gamma_1\gamma_2\gamma_3$. We often omit the
wedge $\wedge$; for discussions of such ``clifform'' notation see
\cite{DMH91,Mi87,Es91}.}
 The covariant differential, $D\Psi:=d\Psi+\omega\Psi$,
includes the Clifford algebra valued connection one-form
$\omega:=\textstyle{1\over4}\gamma_{ab}\omega^{ab}$.
 Now consider the second covariant differential on $\Psi$,
using $D\vartheta^a=0$,
 (i.e., the torsion 2-form vanishes for the Levi-Civita connection),
we obtain
\begin{eqnarray}
 2\, D^2\Psi &=& 2\, \vartheta \wedge D^2\psi =
\textstyle{1\over2}\vartheta^m\wedge\Omega^{ab}\gamma_m\gamma_{ab}\psi
   \nonumber\\
  &=& \vartheta_{[ a} \wedge\Omega^{ab}\gamma_{b ]}\psi
 - \eta_{abc} \wedge\Omega^{ab} \gamma_5 \gamma^{c}\psi .
\end{eqnarray}
Here we have introduced the convenient (Hodge) dual basis
$\eta^{a\dots}:=\ast (\vartheta^a\wedge\cdots)$
and used the identity
 $\gamma_m\gamma_{ab}=2 g_{m [a}\gamma_{b ]}
 - \epsilon_{mabc}\gamma^c\gamma_5$.
The first term vanishes by the first Bianchi
identity and the second term,
\begin{equation}
- \eta_{abc} \wedge\Omega^{ab} \gamma^{c}\psi
  =  - G^b{}_a \gamma_5 \gamma^a \eta_{b} \psi,
\end{equation}
is proportional to the Einstein tensor. This provides a succinct
representation of the Einstein equation.

The Quadratic Spinor Lagrangian (QSL)
\cite{NT95,TJ95,Robinson96,Robinson98,Wallner} is given by
\begin{equation}
 S[\Psi, \omega^{ab}]=
 \int{\cal L}_{\Psi} =
 \int 2 D\overline\Psi \gamma_5 D\Psi.
\end{equation}
 This QSL satisfies the spinor-curvature identity \cite{NTZ}
\begin{equation}
{\cal L}_{\Psi}=2D\overline\Psi \gamma_5D\Psi \equiv
2\overline\Psi\Omega\gamma_5\Psi+d[(D\overline\Psi)\gamma_5\Psi+
\overline\Psi\gamma_5D\Psi ], \label{DPsi^2}
\end{equation}
where
$\Omega={1\over4}\Omega^{ab}\gamma_{ab}=d\omega+\omega\omega$, is
the Clifford algebra valued curvature 2-form. The field equations
$D^2\Psi=0$ and $D(\overline{\Psi}\gamma_{ab}\gamma_5\Psi)=0$ are
equivalent to the Einstein equation and torsion free equation,
respectively.  The metric is given by
$g_{\mu\nu}=\overline\Psi_{(\mu} \Psi_{\nu)}$ . The rhs of
(\ref{DPsi^2}) expands to
\begin{equation}
\overline\psi\psi \Omega^{ab}\wedge\eta_{ab}
 +\overline\psi\gamma_5 \psi \Omega_{ab}\wedge
\vartheta^a\wedge\vartheta^b +d[D(\overline\psi\vartheta)\gamma_5
\vartheta\psi+ \overline\psi\vartheta \gamma_5 D(\vartheta\psi)].
\end{equation}
Since $\Omega^{ab}\wedge\eta_{ab}=-R\ast 1$, for a spinor field
$\psi$, normalized according to
\begin{equation}
\overline\psi\psi=1,\qquad \overline\psi\gamma_5\psi=0,
\label{norms}
\end{equation}
we find that this QSL differs from the standard Hilbert scalar
curvature Lagrangian  only by an exact differential,
\begin{eqnarray}
{\cal L}_\psi&=&2D(\overline{\psi}\vartheta)\gamma_5
D(\vartheta\psi)\nonumber
\\        &\equiv& \Omega^{ab}\wedge\eta_{ab}
        +  d[D(\overline{\psi}\vartheta)\gamma_5\vartheta\psi+
           \overline{\psi}\vartheta\gamma_5 D(\vartheta\psi)].
\label{R+dB}
\end{eqnarray}
In the action this corresponds to a boundary term which does not
affect the local equations of motion.

\section*{Spinor Gauge Invariance of the QSL}

{} From the form of the Lagrangian (\ref{R+dB}), the QSL action
for an extended region actually depends on the (normalized) spinor
field only through the boundary term, not locally.  A change of
the spinor field within the interior of the region will leave the
action unchanged. Consequently the Dirac spinor field $\psi$ has
complete local gauge invariance subject to the two restrictions
(\ref{norms}). This six real parameter spinor gauge freedom can be
represented in the form $\psi=U\psi_0$ where $\psi_0$ is a
normalized Dirac spinor with constant components and $U$ is the
Dirac spinor representation of a Lorentz transformation.  Thus the
gauge freedom of the normalized spinor field is a kind of local
Lorentz gauge freedom.  Considering the scalar curvature term in
the Lagrangian (\ref{R+dB}), it can be recognized that the theory
also has the usual local Lorentz gauge freedom associated with
transformations of the orthonormal frame. Hence there appears to
be two Lorentz gauge freedoms here. But are they really
independent?

The boundary term is
 \begin{eqnarray}
(D\overline{\psi}\gamma_a\gamma_5\gamma_b\psi-
           \overline{\psi}\gamma_a\gamma_5\gamma_b D\psi)
\vartheta^a\wedge\vartheta^b && \nonumber\\
+(\overline{\psi}\gamma_a\gamma_5\gamma_b\psi-
           \overline{\psi}\gamma_b\gamma_5\gamma_a \psi)
D\vartheta^a\wedge\vartheta^b .&&
\end{eqnarray}
Let us consider a gauge transformed spinor field $\psi'=U\psi$.
Then $\overline{\psi'}=\overline{\psi}U^{-1}$, $D\psi'=UD\psi$ and
$D\overline{\psi'}=D(\overline{\psi})U^{-1}$. The gauge
transformed boundary term then becomes
\begin{eqnarray}
(D\overline{\psi}U^{-1}\gamma_a U\gamma_5 U^{-1}\gamma_b U\psi-
  \overline{\psi}U^{-1}\gamma_a U\gamma_5 U^{-1}\gamma_b UD\psi)
\vartheta^a\wedge\vartheta^b &&\nonumber\\
+(\overline{\psi}U^{-1}\gamma_a U\gamma_5 U^{-1}\gamma_b U\psi-
  \overline{\psi}U^{-1}\gamma_b U\gamma_5 U^{-1}\gamma_a U\psi)
D\vartheta^a\wedge\vartheta^b. &&
\end{eqnarray}
The unitary transformations on the gammas induce Lorentz
transformations, $U^{-1}\gamma_a U=\gamma_c L^c{}_a$, on the
orthonormal frame indices. Thus, the six parameter spinor gauge
freedom $\psi$ (with normalization condition) is entirely
equivalent to applying the transformation
$\vartheta'{}^c=L^c{}_a\vartheta^a$ to the orthonormal frame
alone.  Hence the boundary term really has one physically
independent Lorentz gauge freedom.

We showed that the QSL is dynamically equivalent to the tetrad
(teleparallel) representation \cite{TN99}. In the QSL we have a
spinor field which has a six parameter local gauge freedom which
effectively replaces the local Lorentz frame gauge freedom of the
tetrad representation.

\section*{Gravitational Energy-Momentum Density}

{} From Noether's theorem, with $ {\cal E}_\mu=\Sigma^\sigma{}_\mu
\ast dx_\sigma ={\partial_\mu} \rfloor d\phi\wedge (\partial{\cal
L}/\partial d\phi) -{\partial_\mu}\rfloor {\cal L}(\phi)$, we
obtain a canonical energy-momentum 3-form,
\begin{equation}
{\cal E}_\mu
= D \overline\Psi_\mu \wedge \gamma_5 D\Psi
       + D\overline\Psi \wedge \gamma_5 D \Psi_\mu,
\label{Emu}
\end{equation}
satisfying the conservation of energy-momentum $ d{\cal E}_\mu=0$.
Here the covariant differential $D$ only operates on the spinor
but not on the spacetime index.  Expression (\ref{Emu}) is still
another pseudotensor, however, if we consider $\Psi$ as a field in
Minkowski spacetime, then this is a gravitational energy momentum
{\it tensor}. Therefore classically, whether we need the concept
of an energy-momentum density or not, depends on our viewpoint:
(i) When treating the spacetime to be curved, we have no spacetime
symmetry in general. According to the equivalence principle, there
is no well defined energy-momentum density. (ii) With a Minkowski
line element, associated with the spacetime symmetry we have a
covariant energy-momentum density. This is useful in microscopic
physics, where space and time lose its operational meaning.

\section*{Discussion}

By changing the ten parameter metric variables $g_{\mu\nu}$ to the
sixteen parameter tetrad variables $e^a{}_\mu$ in the variational
principle, M{\o}ller \cite{Mo61,Goldberg} obtained a local
energy-momentum density which is a tensor wrt coordinate
transformations but is not a tensor wrt local Lorentz frame
rotations. Recently de~Andrade, Guillen and Pereira
\cite{Pereira00} gave a refined version of M{\o}ller's expression,
but it still depends on a local Lorentz frame rotation. In this
paper, we showed that by adding an auxiliary Dirac spinor field to
the action, we can obtain an expression which is a tensor wrt both
coordinate and local Lorentz frame rotations. This extra six
parameter spinor gauge freedom (eight parameter Dirac spinor with
two normalization conditions) was shown to be equivalent to the
Lorentz transformation for the associated orthonormal frame.

By comparing the formulation with Yang-Mills gauge theory ({\it
Table \ref{table1}}), we find that we can define a gravitational
field strength $F_{\rm G}=D\Psi$.

\begin{table}[ht]
\caption{Comparison between Quadratic Spinor Representation of
General Relativity and Yang-Mills Gauge Theory} \label{table1}
\begin{tabular*}{\textwidth}{@{\extracolsep{\fill}}lcr}
\hline  & \it Yang-Mills     & \it Quadratic Spinor GR \\
\hline &&\\
Potential  &$A=A^I{}_\mu dx^\mu T_I$
&$\Psi=e^I{}_\mu dx^\mu\gamma_I\psi$\\
&&\\ Field Strength  &$F_{\rm YM}=DA$ &$F_{\rm G}=D\Psi$ \\
&&\\ Lagrangian &${\cal L}_{\rm YM}=tr F\wedge\ast F$&${\cal
L}_{\rm QS}=\overline{F_{\rm G}} \gamma_5 F_{\rm G}$ \\
&&\\ Field equations
  & $D\ast F=0$            &  $ D\gamma_5 F_{\rm G}=0$ \\ \hline
\end{tabular*}
\end{table}

Presently it is not so clear here what is the corresponding gauge
group in this formulation. Several related approaches may clarify
the situation: (1) using a semi-direct sum of the group $SL(2,C)$
and $C^4$ \cite{Robinson98}, (2) the Teleparallel approach
\cite{GronHehl,TN99,Cho76,AP97}, (3) considering the spinor
one-form $\Psi$ as an anticommuting field
\cite{BarM,MM,Mansouri,Tung00}. We also note that there is an
interesting generalization of the quadratic  spinor Lagrangian to
the Einstein-Maxwell system \cite{Robinson00}.

We close by noting that the approach discussed here also suggests
that we might be able to find an expression for a covariant
gravitational ``Lorentz force law''.

\end{document}